\documentclass{iau} 
\usepackage{graphicx}

\title[Chemical abundances in GCs] 
{Chemical abundances of multiple stellar populations in massive globular clusters}

\author[Anna F. Marino]   
{Anna F. Marino$^1$}

\affiliation{$^1$Research School of Astronomy \& Astrophysics, Australian National University, \\ Mt Stromlo Observatory, via Cotter Rd, Weston, ACT 2611, Australia \\
email: {\tt anna.marino@anu.edu.au} }

\pubyear{2015}
\volume{316}  
\jname{Formation, evolution, and survival of massive star clusters}
\jname{Formation, evolution, and survival of massive star clusters}
\editors{A.C. Editor, B.D. Editor \& C.E. Editor, eds.}
\begin{document}

\maketitle

\begin{abstract}
Multiple stellar populations in the Milky Way globular clusters manifest themselves with a large variety. 
Although chemical abundance variations in light elements, including He, are ubiquitous, the amount of these variations is different in different globulars. 
Stellar populations with distinct Fe, C+N+O and slow-neutron capture elements have been now detected in some globular clusters, whose number will likely increase. All these chemical features correspond to specific photometric patterns.
I review the chemical+photometric features of the multiple stellar populations in globular clusters and discuss how the interpretation of data is being more and more challenging. Very excitingly, the origin and evolution of globular clusters is being a complex puzzle to compose.


\end{abstract}

\firstsection

\section{Introduction}
The modern picture of globular clusters (GCs) is different from the idea of these objects being a simple stellar population.
Thanks to high precision photometry we know that the color-magnitude diagram (CMD) of all investigated Milky Way GCs exhibit multiple photometric sequences along all evolutionary stages, from the main sequence (MS) up to the red giant branch (RGB) and the horizontal branch (HB) (e.g.\,\cite[Milone et al.\,2012]{Milone2012}; \cite[Piotto et al.\,2015]{Piotto2015}).

Multiple photometric sequences are the effect of the presence of different stellar populations, with their distinct chemical composition and stellar internal structure (e.g.\,\cite[Marino et al.\,2008, 2012a]{Marino2008,Marino2012a}).
The list of the most investigated chemical elements that have been observed to vary among stars in a given GC includes: He, Li, C, N, O, C+N+O, Na, Mg, Al, Si, K, Fe, and many neutron-capture ($n$-capture) elements. 
Not all GCs exhibit the same chemical pattern, and most of them do not show evidence for significant abundance variations in many of these elements.
In the same way, each photometric split along different evolutionary stages of the CMD tells us a different story about the nature of the multiple stellar populations hosted in GCs.

I review the chemical abundance variations observed in the Milky Way GCs and the way we identify them photometrically. I will focus on the more massive GCs, that are those exhibiting more complex and puzzling chemical/photometric patterns. 

\section{Light elements variations}

The most common observed star-to-star abundance patterns in Galactic GCs (GGCs) are the C-N and O-Na anticorrelations. Chemical variations in CNONa have been found to be ubiquitous among the GGCs investigated so far (e.g.\,\cite[Carretta et al.\,2009]{Carretta2009}).
On the other hand, the size of these variations appears to change with the mass, as more massive GCs are generally those having the more extended anticorrelations.

\begin{figure}[b]
\begin{center}
 \includegraphics[width=5.3in]{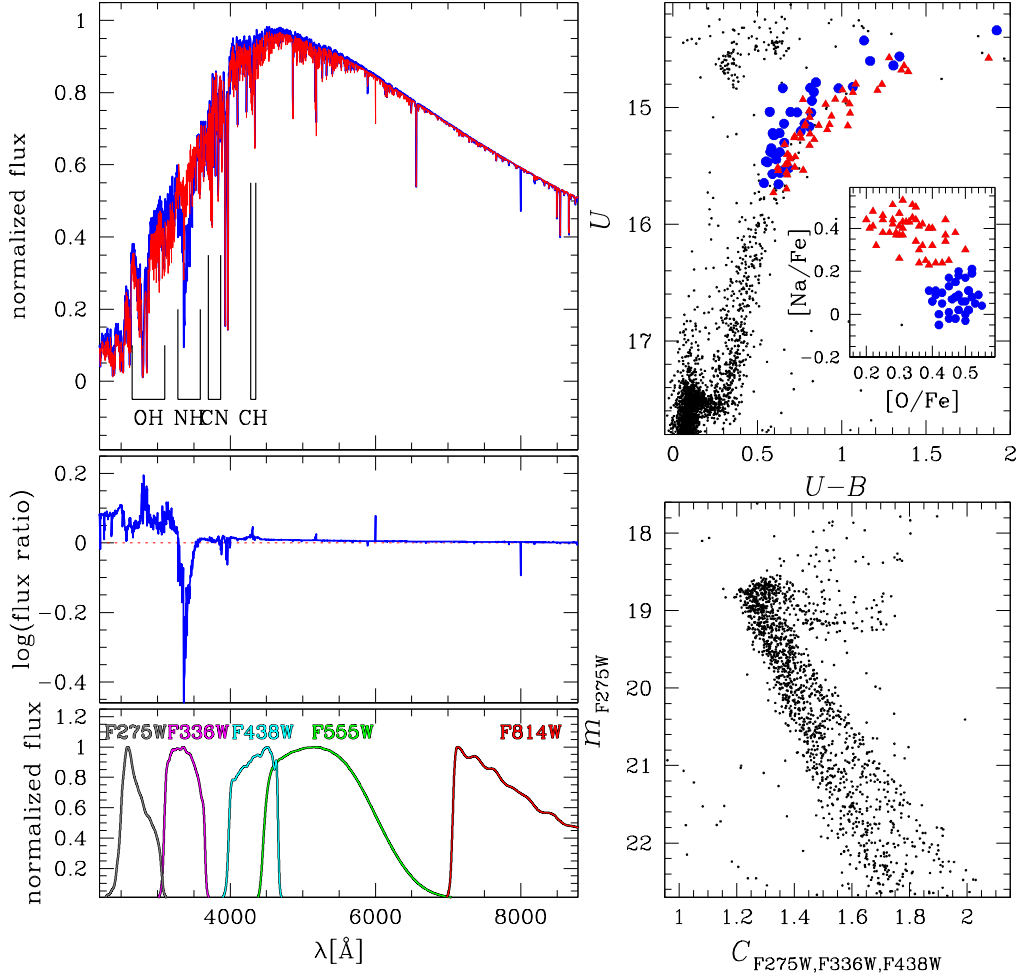} 
 \caption{{\it Right panels:} The $U$-$(U-B)$ CMD from ground-based photometry for M\,4, with the location of N-Na-poor/C-O-rich and N-Na-rich/C-O-poor stars (\cite[Marino et al.\,2008]{Marino2008}) (upper panel). The lower panel represents the {\it HST} $m_{F275W}$-$c_{F275W,F336W,F438W}$ diagram around the MS of M\,4 (\cite[Piotto et al.\,2015]{Piotto2015}). Both the RGB (upper panel) and MS (lower panel) splits in these bands are due to light elements variations. {\it Left panels:} Synthetic spectra for RGB stars computed with the mean C/N/O abundances of the first (blue) and second (red) population in M\,4, with the location of the relevant molecular bands (upper panel). The middle panel displays the difference between the two spectra. The bandpasses of the {\it HST} filters is shown in the lower panel. The {\it HST} Survey of multiple populations observes GCs in filters $F275W$, $F336W$ and $F438W$. }
   \label{fig1}
\end{center}
\end{figure}

Chemical variations in light elements are responsible for many of the multiple sequences observed along the CMDs.
As displayed in Fig.~\ref{fig1}, carbon, nitrogen and oxygen form strong CN, NH, CH, and OH molecular bands in the ultraviolet region of the spectrum and cause the split RGBs and MSs observed in CMDs constructed by using the UV bands. 
A clear example of this behaviour is M\,4, represented in Fig.~\ref{fig1}, which clearly exhibits two RGBs populated by C-O-rich/N-Na-poor and C-O-poor/N-Na-rich stars. Indeed, stars with different content in these elements define different sequences along the CMD when the $(U-B)$ color is used (e.g.\,\cite[Marino et al.\,2008]{Marino2008}). 
The huge effort, through the Legacy GO-13297 (\cite[Piotto et al.\,2015]{Piotto2015}), to analyse the GCs multiple stellar populations in UV bands with \textit{HST} is based on this finding.

Helium variations have been observed within GCs. 
Multi-wavelength photometry of GCs has been often used to infer the relative helium abundances in GCs. Multiple MSs and RGBs, detected in most clusters (\cite[Milone\,2015]{Milone2015}), are interpreted as due to the presence of stellar populations with different He content (e.g.\,\cite[Norris\,2004]{Norris2004}; \cite[Piotto et al.\,2005]{Piotto2005}). 
A few GCs have highly He-enhanced second populations. Among the GCs with highest He enhancements one of the most studied is NGC2\,2808. Photometry suggests the presence of at least five distinct stellar populations that span a wide range of He abundance, up to $Y=0.40$ (\cite[Milone et al.\,2015]{Miloneetal2015}).
The degree of internal variations in He correlates with the mass of the cluster as shown in \cite[Milone\,(2015)]{Milone2015}.

On the spectroscopic side, the determination of He contents is difficult because IR He lines are subject to chromospheric effects and lines in the optical can only be observed at high temperatures. To date, evidence for the presence of significant He variations using spectroscopic analysis have been provided for NGC\,2808 and $\omega$~Centauri (\cite[Dupree et al.\,2013]{Dupree2013}; \cite[Pasquini et al.\,2011]{Pasquini2011}). For NGC\,2808, high He abundances have been inferred for blue HB stars (\cite[Marino et al.\,2014a]{Marino2014a}). 

Some GCs also exhibit Mg-Al anticorrelations (e.g.\,\cite[Norris \& Da Costa\,1995]{NorrisDaCosta1995}; \cite[Yong et al.\,2003]{Yong2003}; \cite[Sneden et al.\,2004]{Sneden2004}; \cite[Carretta\,2015]{Carretta2015}). In many GCs only variations in Al are observed, while Mg does not change significantly.

\section{Anomalous GCs}\label{sec:anomalous}

It is a long time that high resolution spectroscopy revealed large internal variations in Fe and in the elements produced via slow $n$-capture reactions ($s$-elements) in the most massive GGC $\omega$~Centauri (e.g.\,\cite[Norris \& Da Costa\,1995]{NorrisDaCosta1995}; \cite[Smith et al.\,2000]{Smith2000}). 
More recently we realised that, although this phenomenon is quite rare in Milky Way GCs, it is not confined to $\omega$~Centauri.
Indeed, a few GCs (``anomalous") have been found to show a genuine internal variation in the bulk metallicity (e.g.\,\cite[Marino et al.\,2009]{Marino2009}, \cite[Da Costa et al.\,2009]{DaCosta2009}; \cite[Ferraro et al.\,2009]{Ferraro2009}; \cite[Carretta et al.\,2011]{Carretta2011}, \cite[Yong et al.\,2014]{Yong2014}; \cite[Marino et al.\,2015]{Marino2015}; \cite[Johnson et al.\,2015]{Johnson2015}). 

Some of the {\it anomalous} GCs, specifically NGC\,1851, M\,22, M\,2, NGC\,5286 and M\,19 exhibit internal variations in $s$-process elements (\cite[Yong \& Grundahl\,2008]{Yong2008}; \cite[Marino et al.\,2009, 2011a]{Marino2009}; \cite[Yong et al.\,2015]{Yong2015}; \cite[Marino et al.\,2015]{Marino2015}; \cite[Johnson et al.\,2015]{Johnson2015}); and for NGC\,1851 and M\,22 significant variations in the total C+N+O have been also detected (\cite[Yong et al.\,2015]{Yong2015}; \cite[Marino et al.\,2011a]{Marino2011a}).
The $s$-elements and C+N+O abundances are positively correlated to Fe variations; the common light elements variations, typical of {\it normal} GGCs, are independently present within each population with different Fe/$s$-elements/C+N+O (\cite[Marino et al.\,2009, 2011a]{Marino2009}).
Both the $s$-elements and the total C+N+O chemical contents are instead uniform in typical GGCs. 

Photometrically, the most striking feature of {\it anomalous} GCs can be identified in the SGB region, which is split (\cite[Milone et al.\,2008]{Milone2008}; \cite[Piotto et al.\,2012]{Piotto2012}). This feature is also peculiar of {\it anomalous} GCs, since typical GGCs exhibit a single SGB in visual bands. 
On theoretical background split SGBs are the indication of stellar populations with different C+N+O or different age (\cite[Cassisi et al.\,2008]{Cassisi2008}).
Spectroscopic+photometric campaigns have confirmed that the SGB splits are produced by stars with different interior structures due to different overall C+N+O and metallicities (\cite[Marino et al\,2012a]{Marino2012a}; \cite[Gratton et al.\,2012]{Gratton2012}; \cite[Marino et al.\,2014b]{Marino2014b}). The RGB of {\it anomalous} GCs is also peculiar, as they show well-separated multiple branches in bands where typical GGCs exhibit single sequences, e.g. in the index $c_{BVI}$ and in the $U$-$(U-V)$, where the faint and the bright SGB are clearly connected with the red and the blue RGB, respectively (\cite[Marino et al.\,2015]{Marino2015}).

\begin{figure}[b]
\begin{center}
 \includegraphics[width=5.1in]{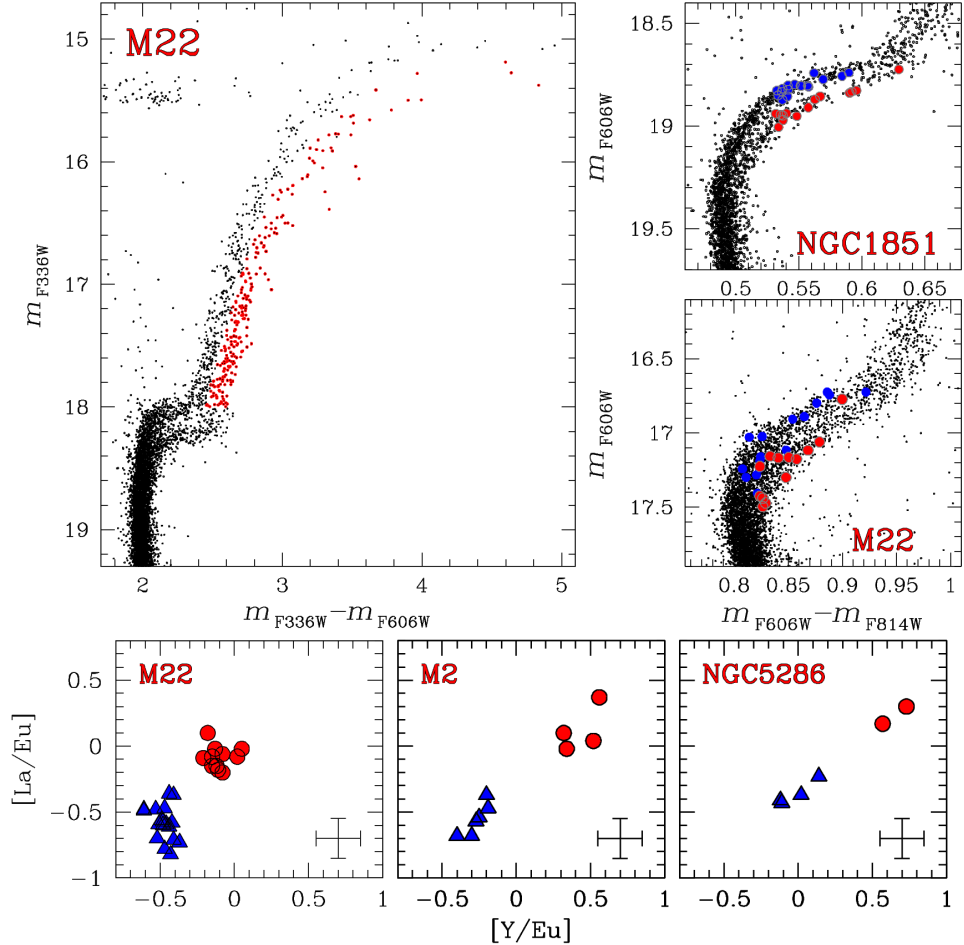} 
 \caption{{\it Upper panels:} CMDs obtained from {\it HST} photometry for the {\it anomalous} GCs M\,22 and NGC\,1851. On the left panel the $m_{F336W}$-$(m_{F336W}-m_{F606W})$ for M\,22 clearly shows the SGB split. The {\it anomalous} faint SGB is connected with a redder RGB sequence (red). The right panels display the $m_{F606W}$-$(m_{F606W}-m_{F814W})$ CMD for NGC\,1851 and M\,22 around the SGB region. In both clusters $s$-poor and $s$-rich stars, as inferred from spectroscopy (\cite[Marino et al.\,2012a]{Marino2012a}; \cite[Marino et al.\,2014b]{Marino2014b}), have been represented in blue and red, respectively. 
{\it Lower panels:} [La/Eu] as a function of [Y/Eu] for the {\it anomalous} GCs M\,22, M\,2 and NGC\,5286 (\cite[Marino et al.\,2009, 2011a]{Marino2011a}; \cite[Yong et al.\,2009]{Yong2009}; \cite[Marino et al.\,2015]{Marino2015}). $s$-poor and $s$-rich stars have been represented with blue triangles and red dots, respectively.
}
   \label{fig2}
\end{center}
\end{figure}

A summary of some spectroscopic and photometric features of {\it anomalous} GCs is presented in Fig.~\ref{fig2}, where the double SGB and RGB of M\,22 due to the C+N+O variations are clearly visible. Chemical abundances for stars distributed on the split SGB has been provided for M\,22 and NGC\,1851, demonstrating that in both cases the $s$-rich stars distribute on the {\it anomalous} faint SGB (see CMDs on the right panels of Fig.~\ref{fig2}). Some chemical variations in $s$-elements for M\,22, M\,2 and NGC\,5286 have been represented in the lower panels of Fig.~\ref{fig2}.

The similarity of these objects with $\omega$~Centauri is remarkable, not just because of the metallicity variations but also for the chemical pattern is $s$-elements. \cite[Da Costa \& Marino\,(2011)]{DaCostaMarino} presented a comparison between M\,22 and $\omega$~Centauri and both the GCs have a very similar rise of the $s$-elements content as a function of Fe in the common Fe-range ([Fe/H]$\lesssim-$1.5, see right panels in Fig.~\ref{fig3}). Additionally both these clusters present similar patterns in the Na-O anticorrelation present among stars with different Fe and C+N+O variations (\cite[Marino et al.\,2011b, 2012b]{Marino2011b,Marino2012b}).

\subsection{On the Fe variations of M\,22}

Recent claim has been made that the {\it anomalous} GC M\,22 does not host stellar populations with different Fe (\cite[Mucciarelli et al.\,2015]{Mucciarelli2015}). 
The argument is that, if photometric gravities are used instead of those from the ionisation equilibrium, the difference in Fe~\,{\sc ii} between the two stellar populations with different $s$/C+N+O content disappears.
The Fe~\,{\sc i} difference instead cannot be removed whatever technique is used to derive the atmospheric parameters. The reason supplied by \cite[Mucciarelli et al.\,(2015)]{Mucciarelli2015} for this behaviour was the presence of non-local thermodynamic effects (NLTE) affecting Fe~\,{\sc i}.

\begin{figure}[b]
\begin{center}
 \includegraphics[width=5.3in]{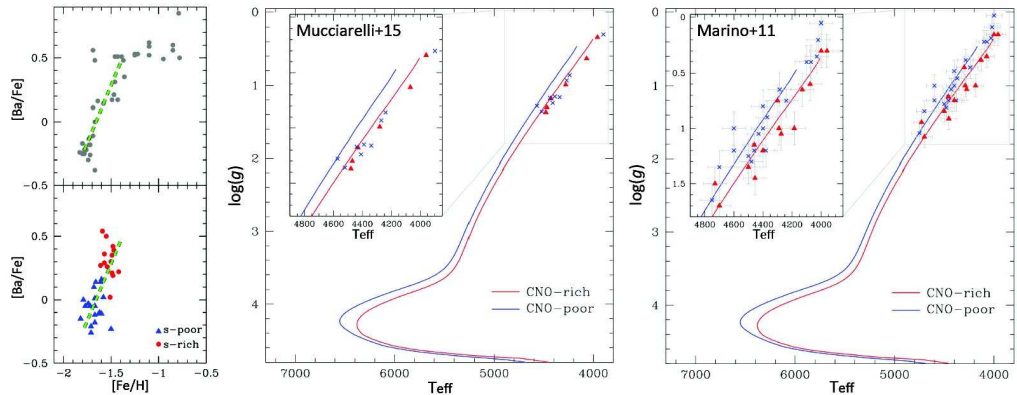} 
 \caption{{\it Left panels}: [Ba/Fe] as a function of [Fe/H] for $\omega$~Centauri (\cite[Norris \& Da Costa\,1995]{NorrisDaCosta1995}) and M\,22 (\cite[Marino et al.\,2009, 2011a]{Marino2009}; shifted by 0.15~dex in [Fe/H]). For M\,22 the $s$-poor and $s$-rich stars have been plotted as blue triangles and red dots, respectively. A straight line (green dashed line) tracing the rise of $s$-elements with Fe for stars with [Fe/H]$\lesssim -$1.5 in $\omega$~Centauri is super-imposed to the M\,22 data. If there are no Fe variations in M\,22, they will also disappear in $\omega$~Centauri for stars with  [Fe/H]$\lesssim -$1.5~dex, removing the well-known $s$-pattern in this GC. 
{\it Right panels}: Theoretical isochrones from BaSTI (\cite[Pietrinferni et al.\,2006]{Pietrinferni2006}) for a CNO-poor and CNO-rich population. Superimposed to the isochrones are the atmospheric parameters for M\,22 RGBs from \cite[Mucciarelli et al.\,(2015)]{Mucciarelli2015} (middle-panel) and \cite[Marino et al.\,(2009, 2011a)]{Marino2011a} (right-panel). The atmspheric parameters from \cite[Mucciarelli et al.\,(2015)]{Mucciarelli2015} force the stars on a single sequence.
} 
   \label{fig3}
\end{center}
\end{figure}

Such an effect, if real, would strongly affect what we know about {\it anomalous} GCs, including $\omega$~Centauri. Indeed, the well-known rise in $s$-process elements with Fe in $\omega$~Centauri lies on the same Fe-range spanned by M\,22 stars. Given the similar Fe and chemical patterns of these two GCs, if the common Fe-regime is considered, there is no reason why the same effects should not apply to $\omega$~Centauri (see left panels of Fig~\ref{fig3}). This would imply that there is no rise of $s$-elements abundances with Fe also in $\omega$~Centauri.

As it will be presented in \cite[Marino et al.\,(in prep)]{Marino_inprep}, however, uncounted NLTE effects on Fe~\,{\sc i} are not responsible for the appearance of the Fe spread in M\,22 (see \cite[Lind et al.\,2012]{Lind2012} for discussion of NLTE effects on Fe).
This result was already suggested by the variations in CaT lines found in M\,22 (\cite[Da Costa et al.\,2009]{DaCosta2009}), that should not be affected by the same NLTE effects.
On the other hand, the disappearance of the Fe~\,{\sc ii} difference in the Mucciarelli et al. analysis is due to the fact that they neglect the C+N+O variations in M\,22 discussed in Sect.~\ref{sec:anomalous} and get significant systematic offset in gravities, that affect {\it only} the $s$-poor population. 
As shown in Fig.~\ref{fig3}, this offset makes the two populations of M\,22 lying on the same isochrone, which is unrealistic given their different C+N+O content, as suggested both from spectroscopy and photometry.
Such systematic in surface gravity affects the abundances obtained from ionised Fe, but does not affect Fe~{\sc i} significantly.

On the other hand, the main advantage of stellar parameters independent on photometry is that they do not introduce systematic errors that affect one stellar population only. This is suggested by the temperature-gravity distribution obtained from \cite[Marino et al.\,(2009, 2011a)]{Marino2011a}, which agrees with that expected from two stellar populations with different C+N+O.

Thus, the Fe variations in M\,22 and its similarity with $\omega$~Centauri are confirmed. There is no support for NLTE affecting this result, but systematics on the atmospheric parameters due to the neglecting of C+N+O variations may introduce spurious results.

\section{The chemical enrichment in GCs}

Explaining the whole observational scenario of multiple stellar populations in GCs is difficult. 
Simply based on the chemical abundances, we may think that {\it normal} and {\it anomalous} GCs have experienced a different chemical evolution. 

\begin{figure}[!b]
\begin{center}
 \includegraphics[width=5.in]{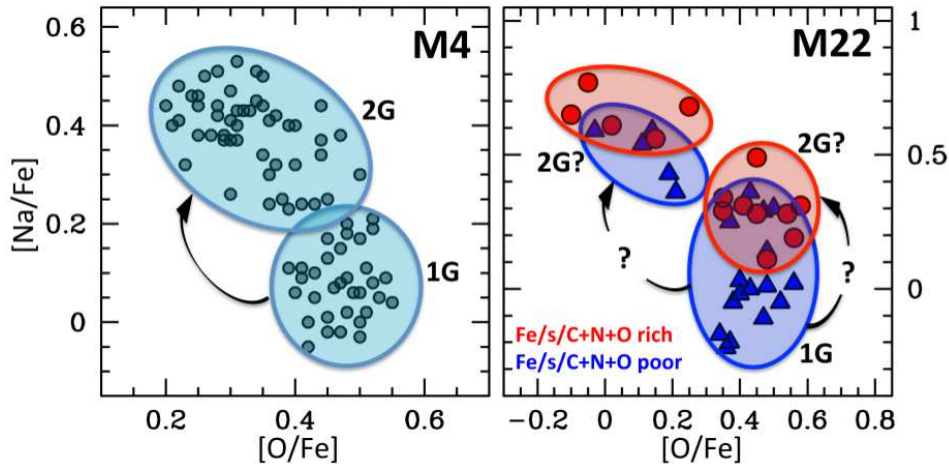} 
 \caption{[Na/Fe] versus [O/Fe] for M\,4 (\cite[Marino et al.\,2008]{Marino2008}) and M\,22 (Marino et al. 2009, 2011a). For M\,4 a two-generation model is sufficient with the second generation (2G), being Na-rich and O-poor as it formed from the first generation (1G) polluters.
For the {\it anomalous} GC M\,22, the observational scenario is more complex: the Na and O variations are present in both the Fe/$s$-elements/C+N+O-poor (blue) and Fe/$s$-elements/C+N+O rich (red) populations.
We do not know yet which is the 2G that formed directly after the 1G, e.g. which enrichment occurred first (in Fe or in hot H-burning products). Similar challenges affect the scenario of massive interacting binaries.}
   \label{fig4}
\end{center}
\end{figure}

Although we know from recent photometric results that GCs host more than two stellar populations (\cite[Milone et al.\,2015]{Milone2015}), a ``two-populations'' scenario can still approximatively explain the light elements chemical patterns observed in {\it typical} GCs, as sketched in Fig.~\ref{fig4} (left panel).
We may assume a self-pollution scenario, e.g. the first generation (1G) is enhanced in O, depleted in Na (just as typical of Galactic halo stars), and, later on, a second generation (2G) forms.
If we suppose a similar scenario, 2G stars form from hot-H burning processed material, which has been realised from some kind of polluter. The proposed candidate polluters are fast-rotating massive stars (\cite[Decressin et al.\,2007]{Decressin2007}), asymptotic giant branch stars (\cite[D'Ercole et al.\,2008]{Dercole2008}), or supermassive stars (\cite[Denissenkov et al.\,2015]{Denissenkov2015}).
If we instead do not invoke multiple star-formation episodes, we have to find other mechanisms that can account for the chemical variations. In this context, massive interacting binaries have been proposed to be able of reproduce some of the chemical variations (\cite[Bastian et al.\,2013]{Bastian2013}).
None of these scenarios at the moment is accepted, as all of them have serious shortcomings in explaining the observations (see \cite[Renzini et al.\,2015]{Renzini2015}). 
 
If explaining {\it normal} GCs is complex, {\it anomalous} GCs present even more challenges. The chemical pattern displayed by these objects is complex and we do not have, at the moment, a solution to interpret the observations in terms of a reasonable chemical enrichment history. As an example, the right panel of Fig.~\ref{fig4} shows the Na and O abundances in M\,22. In this GC internal variations in Na and O exist in stars with different Fe/$s$-elements/C+N+O content. This means that, for this object, as well as in the other {\it anomalous} GCs, we do not know how the intra-cluster chemical enrichment proceeded, e.g. if the first channel to be active was the enrichment in the hot H-burning products (producing light elements patterns) or the enrichment in Fe. Surely, by assuming the self-pollution scenario, polluters of different mass have contributed to the intra-cluster pollution, including low-mass AGBs that produce $s$-elements and increase the C+N+O (e.g.\,\cite[Shingles et al.\,2014]{Shingles2014}; \cite[Straniero et al.\,2014]{Straniero2014}).

The enrichment in Fe suggests that {\it anomalous} GCs have been able of retaining Supernovae ejecta. Metallicity variation, which was thought to be an exclusive feature of $\omega$~Centauri, proposed to be the remnant of a dwarf galaxy (e.g.\,\cite[Bekki \& Freeman\,2003]{BekkiFreeman}), is a more widespread phenomenon in GCs. Thus, it is tempting to speculate that the {\it anomalous} GCs may be nuclei of disrupted dwarf galaxies, as suggested for $\omega$~Centauri.

\section{Conclusive remarks}

Thanks to the large observational material, in the last decade the formation and evolution of GCs has became more difficult to understand.
We do not have any model which is able to satisfy all the observational constraints. I conclude by listing some relevant open issues that we should address in the future: 
\begin{itemize}
\item{in general, the origin of the multiple stellar populations in GCs. Will one or more of the proposed scenarios (self-pollution, fast-rotating massive stars, supermassive stars, mass exchange in massive interacting binaries) able to explain the whole observational scenario?}
\item{could the heterogeneity of the multiple population zoo be reconciled with a unique scenario?}
\item{is there some GCs' property, such as mass, determining the length of star formation, and hence the level of chemical enrichment?}
\item{which is the origin of anomalous GCs? Why they exhibit different chemical enrichment with respect to the majority of GGCs? Does their peculiar chemical pattern simply imply a longer star-formation history? Or, could have they originated as nuclei of disrupted galaxies?}
\end{itemize}

\begin{acknowledgments}
I acknowledge the organisers of the Symposium for inviting me to review results on the multiple stellar populations in globular clusters.
I am grateful to all the collaborators who contribute to the presented results.
\end{acknowledgments}

\end{document}